\documentclass{JHEP3}
\usepackage{graphicx}
\usepackage{amsmath, amsthm, amsfonts}
\newcommand{\beq}{\begin{equation}}
\newcommand{\bea}{\begin{eqnarray}}
\newcommand{\eeq}{\end{equation}}
\newcommand{\eea}{\end{eqnarray}}

\newcommand{\bfg}{\begin{figure}[!htb]}
\newcommand{\efg}{\end{figure}}
\newcommand{\bc}{\begin{center}}
\newcommand{\ec}{\end{center}}

\newcommand{\p}{\partial}
\newcommand{\tphi}{\tilde{\tilde{\phi}}}

\newcommand{\tgamma}{\tilde{\gamma}}
\newcommand{\tlambda}{\tilde{\lambda}}

\newcommand{\bgamma}{\bar{\gamma}}
\newcommand{\dphi}{\dot{\phi}}
\newcommand{\dtphi}{\dot{\tphi}}
\newcommand{\dr}{\dot{r}}
\newcommand{\cL}{\mathcal{L}}
\newcommand{\cR}{\mathcal{R}}
\newcommand{\cJ}{\mathcal{J}}
\newcommand{\tD}{\tilde{D}}
\newcommand{\cD}{\mathcal{D}}
\newcommand{\ctD}{\mathcal{\tD}}

\title{$\gamma_{i}$-Deformed Lax Pair for Rotating Strings in the Fast Motion Limit}
\author{A.H. Prinsloo\\
Department of Physics and Center for Theoretical Physics, University of the Witwatersrand, Private Bag 3,
Wits, 2050, South Africa\\
Email: prinslooa@students.wits.ac.za}

\abstract{

\par{A 3-parameter generalization of the Lunin-Maldacena
background has recently been constructed by Frolov. This
$\gamma_i$-deformed background is non-supersymmetric. We consider
strings in this $\gamma_{i}$-deformed $\mathbb{R}\times S^{5}$
background rotating in three orthogonal planes (the 3-spin sector)
in a fast motion limit, in which the total angular momentum $J$ is
assumed to be large.  We show that there exists a consistent
transformation which takes the undeformed equations of motion into
the $\gamma_{i}$-deformed equations of motion.  This
transformation is used to construct a Lax pair for the bosonic
part of the $\gamma_{i}$-deformed theory in the fast motion limit.
This implies the integrability of the bosonic part of the
$\gamma_{i}$-deformed string sigma model in the fast motion
limit.} }

\preprint{} 

\begin{document}

\maketitle

\newpage

\section{Introduction}

\par{The original conjecture of Maldacena \cite{malda}, which was further elaborated in \cite{gub-kleb-poly1,wit},
claims a correspondence between string theory in an $AdS_{5}\times S^{5}$ background and $\mathcal{N}=4$ Super
Yang-Mills (SYM) theory. It is well known \cite{leigh-strass} that one can construct a marginal deformation of
 $\mathcal{N}=4$ SYM theory to obtain an $\mathcal{N}=1$ superconformal SYM theory.
These include the so-called $\beta$-deformations. Lunin and Maldacena have found gravity backgrounds which are
conjectured to be dual to these $\beta$-deformations \cite{lunin-malda}. Evidence for this conjecture includes
the matching of the energies of semi-classical strings to the anomalous dimensions of corresponding operators in
the gauge theory \cite{fro-roi-tsey1} and a study of the pp-wave limit of the string theory together with an
identification of the dual BMN operators \cite{niar-prez,kms,mateos}. Furthermore, the central charge was
reproduced in the dual gravitational description and shown to be independent of the deformation parameter
\cite{pal,freed-gursoy}. This verifies on the string theory side that $\beta$-deformations are indeed marginal.
Frolov was able to find a Lax pair representation for the $\beta$-deformed string theory in the case of a real
deformation parameter $\beta = \gamma$ \cite{frolov}. The crucial insight in this work was that the
Lunin-Maldacena deformation, in the case of a real deformation parameter $\gamma$, can be realized as a TsT
transformation of the string action with shift parameter $\tgamma = \sqrt{\lambda}\gamma$, where $\sqrt{\lambda}
= R^{2}$ and $R$ is the radius of $S^{5}$.}

\par{Frolov also constructed a non-supersymmetric $\gamma_{i}$-deformed string theory by performing a
series of three TsT transformations with shift parameters
$\tgamma_{i} = \sqrt{\lambda}\gamma_{i}$ on the string action and
conjectured a correspondence to a non-supersymmetric
$\gamma_{i}$-deformed SYM theory \cite{frolov}\footnote{For
further discussions of backgrounds arising from TsT
transformations see \cite{russo,rash-vis-yang}.}. Remarkably, this
3-parameter deformation also admits a Lax pair representation.
Evidence for this proposed correspondence was given by Frolov,
Roiban and Tseytlin \cite{fro-roi-tsey2}, who matched the
semi-classical energies of strings to the anomalous dimensions of
gauge theory operators using a $\gamma_{i}$-deformed spin chain
representation. Furthermore, agreement was also found between the
string and gauge theory descriptions of open strings attached to
unstable giant gravitons \cite{kns}. These comparisons are
particularly interesting because any agreement found cannot be the
result of supersymmetry on either side, since these
$\gamma_{i}$-deformed string and gauge theories are
non-supersymmetric. The special case of the $\gamma_{i}$-deformed
background with equal shift parameters $\gamma_{i} = \gamma$ is
the same as the $\beta$-deformed background with real $\beta =
\gamma$.}

\par{The conjectured gauge/string theory duality is of the strong-weak coupling type with respect to the 't Hooft
coupling. This makes comparison between the two sides difficult. In this regard, studying a semi-classical limit
of the theory is surprisingly fruitful as it allows perturbative computations in the gauge theory to be matched
to classical computations in the dual gravitational theory \cite{gub-kleb-poly2,fro-tsey}.}

\par{The semi-classical limit formulated and used in \cite{fro-roi-tsey1,fro-roi-tsey2,kruc,dim-rash,
kruc-ryz-tsey,hern-lopez,stef-tsey,kruc-tsey}, which will
henceforth be known as the fast motion limit, will now be
described in detail. We consider strings in $\mathbb{R}\times
S^{5}$ rotating in three orthogonal planes (the 3-spin
sector)\footnote{Multispin solutions describing rotating strings
in the Lunin-Maldacena background were found in
\cite{bob-dim-rash}.} and assume that the total angular momentum
$J$ is large. The string action in the fast motion limit is then
obtained by isolating the ``fast" angular coordinate (which
corresponds to the total angular momentum), choosing a suitable
gauge and assuming that the time derivatives of the radii and
other ``slow" angular coordinates are small (of order $\tlambda =
\frac{\lambda}{J^{2}}$). The gauge generally used is known as the
non-diagonal uniform gauge and ensures that the angular momentum
is spread evenly along the string (this makes comparison to the
effective action of the corresponding spin chain on the gauge
theory side possible). }

\par{The Lagrangian governing the dynamics of strings rotating in three orthogonal planes in the $\gamma_{i}$-deformed
$\mathbb{R}\times S^{5}$ background in the fast motion limit was derived in \cite{fro-roi-tsey2} and shown to be
equivalent on the gauge theory side to the effective action of a $\gamma_{i}$-deformed spin chain in the
continuum limit (where the length of the spin chain $J$ becomes large). Now, for the case of a similar
Lagrangian in the undeformed background (which was originally derived in \cite{hern-lopez, stef-tsey,kruc-tsey},
but was also obtained in \cite{fro-roi-tsey2} by setting $\gamma_{i}$=0), the undeformed equations of motion are
equivalent to a Landau-Lifshitz equation for which there is a known Lax pair representation
\cite{fro-roi-tsey2,stef-tsey}. It is not known, however, whether the dynamics following from the
$\gamma_{i}$-deformed Lagrangian are integrable. In this article we will construct a Lax pair representation for
the $\gamma_{i}$-deformed equations of motion, thereby settling this issue.}

\par{In section 2 we review the undeformed case.  The undeformed equations of motion in the fast motion limit
are derived and shown to be equivalent to the zero curvature condition for the undeformed Lax pair. In section 3
we derive the $\gamma_{i}$-deformed equations of motion and construct a transformation which takes the
undeformed equations of motion into these $\gamma_{i}$-deformed equations of motion.  We then choose a suitable
gauge for the undeformed Lax pair and use this transformation to derive a $\gamma_{i}$-deformed Lax pair. A
brief conclusion is then given in section 4 and some of the more lengthy calculations are included in
appendices.}

\section{Strings in the Undeformed Background}

\subsection{Equations of Motion in the Undeformed Background}

\par{Let us consider strings in the undeformed $\mathbb{R}\times S^{5}$ background (at the center of $AdS_{5}$)
rotating in three orthogonal planes. The total angular momentum is
$J = J_{1}+J_{2}+J_{3}$, where $J_{i}$ is the angular momentum
associated with the $i^{\textrm{th}}$ angular coordinate
$\tphi_{i}$ (where i=1, 2, 3)\footnote{We make use of the notation
of \cite{frolov} for the angular coordinates in the undeformed and
$\gamma_{i}$-deformed backgrounds.}. In the fast motion limit,
where the total angular momentum $J$ is large and the time
derivatives of the radii and ``slow" angular coordinates are
assumed to be of order $\tlambda = \frac{1}{\cJ^{2}} =
\frac{\lambda}{J^{2}}$, the string action to first order in
$\tlambda$ (derived in \cite{fro-roi-tsey2}) is}

\beq S = J\int{d\tau\frac{d\sigma}{2\pi}[\cL + O(\tlambda^{2})]}~,
\eeq

\par{where the Lagrangian in the undeformed background is\footnote{It should be noted that henceforth
the Einstein summation convention will not be used.  All summations will be explicitly mentioned so as to avoid
confusion.}}

\beq \cL = \sum_{i=1}^{3}{r_{i}^{2}\dtphi_{i}} - \frac{\tlambda}{2}\{\sum_{i=1}^{3}(r_{i}')^{2} +
\sum_{^{i,j=1}_{i<j}}^{3}{r_{i}^{2}r_{j}^{2}(\tphi_{i}'-\tphi_{j}')^{2}} + \Lambda (\sum_{i=1}^{3}r_{i}^{2} - 1)
\}~. \eeq

\bigskip

\par{The last term is a constraint term. Re-defining $\tau \rightarrow \frac{1}{\tlambda}\tau$
and $\cL \rightarrow -\frac{1}{\tlambda}\cL$ gives}

\beq \cL = -\sum_{i=1}^{3}{r_{i}^{2}\dtphi_{i}} + \frac{1}{2} \sum_{i=1}^{3}{(r_{i}')^{2}} +
\frac{1}{2}\sum_{^{i,j=1}_{i<j}}^{3}{r_{i}^{2}r_{j}^{2}(\tphi_{i}'-\tphi_{j}')^{2}} + \frac{1}{2}\Lambda
(\sum_{i=1}^{3}r_{i}^{2} - 1)~. \eeq

\bigskip

\par{The equations of motion, obtained by varying with respect to $r_{i}$ and $\tphi_{i}$ respectively, are}

\bea \label{eqn-und-motion1} & r_{i}'' & = -2r_{i}\dtphi_{i} + r_{i}\sum_{k=1}^{3}{r_{k}^{2}(\tphi_{i}'-\tphi_{k}')^{2}} + \Lambda r_{i}~, \\
     \label{eqn-und-motion2} & \dr_{i} & = \sum_{k=1}^{3}r_{k}(r_{i}r_{k})'(\tphi_{i}'-\tphi_{k}') +
                   \frac{1}{2}r_{i}\sum_{k=1}^{3}r_{k}^{2}(\tphi_{i}''-\tphi_{k}'')~, \eea

\par{while varying with respect to the Lagrange multiplier $\Lambda$ gives the constraint
equation $\sum\limits_{i=1}^{3}{r_{i}^{2}} = 1$.}

\bigskip

\par{Now, assuming this constraint is satisfied, the equations of motion (\ref{eqn-und-motion1})
and (\ref{eqn-und-motion2}) are equivalent to}

\bea \label{eqn-und_comm1} & r_{j}r_{i}''-r_{i}r_{j}'' = & 2r_{i}r_{j}(\dtphi_{j}-\dtphi_{i}) +
r_{i}r_{j}\sum_{k=1}^{3}{r_{k}^{2}(\tphi_{i}'-\tphi_{k}')^{2}} -
r_{i}r_{j}\sum_{k=1}^{3}{r_{k}^{2}(\tphi_{j}'-\tphi_{k}')^{2}}~, \\
\nonumber && \\
\nonumber & & \\
\label{eqn-und_comm2} \nonumber & \dr_{i}r_{j} + r_{i}\dr_{j} = &
r_{j}\sum_{k=1}^{3}{r_{k}(r_{i}r_{k})'(\tphi_{i}'-\tphi_{k}')} +
r_{i}\sum_{k=1}^{3}{r_{k}(r_{j}r_{k})'(\tphi_{j}'-\tphi_{k}')} \\
& & +
\frac{1}{2}r_{i}r_{j}\sum_{k=1}^{3}{r_{k}^{2}(\tphi_{i}''-\tphi_{k}'')}+
\frac{1}{2}r_{i}r_{j}\sum_{k=1}^{3}{r_{k}^{2}(\tphi_{j}''-\tphi_{k}'')}~.\eea

\par{Notice that the constraint term cancels out of equation (\ref{eqn-und_comm1}).}

\subsection{Lax Pair in the Undeformed Background}

\par{A Lax Pair for this undeformed system \cite{fro-roi-tsey2}, which is a function of
the spectral parameter $x$, is }

\beq \cD_{\mu} = \p_{\mu} - A_{\mu} ~~~~~~~~ \textrm{with} ~~ \mu = 0,1 ~, \eeq

\par{where}

\bea & A_{0} & = \frac{1}{6}[N,\p_{1}N]x+\frac{3i}{2}Nx^{2}~, \\
     & A_{1} & = iNx \eea

\par{and we have defined $N_{ij} = 3U_{i}^{*}U_{j} - \delta_{ij}$, where $U_{i} = r_{i}e^{i\tphi_{i}}$ and
$\sum\limits_{i=1}^{3}{r_{i}^{2}}=1$. }

\bigskip

\par{This satisfies the zero curvature condition $[\cD_{0}, \cD_{1}] = 0$,
which is equivalent to}

\beq \label{eqn-undeformed lax} \p_{0}A_{1} - \p_{1}A_{0} -
[A_{0},A_{1}] = 0~. \eeq

\bigskip

\par{The above equation results in the Landau-Lifshitz equation of motion
\cite{fro-roi-tsey2,stef-tsey} given by}

\beq i\p_{0}N = \frac{1}{6}[N,\p_{1}^{2}N] \eeq

\par{and, upon substitution of $N_{ij} = 3U_{i}^{*}U_{j}-\delta_{ij} = 3r_{i}r_{j}e^{i(\tphi_{j}-\tphi_{i})}-\delta_{ij}$
into this equation, one obtains equations (\ref{eqn-und_comm1})
and (\ref{eqn-und_comm2}), which are equivalent to the undeformed
equations of motion (see appendix A).}

\bigskip

\par{In terms of $r_{i}$ and $\tphi_{i}$, the undeformed Lax pair is $\cD_{\mu} = \p_{\mu} - A_{\mu}$, where}

\beq \label{eqn-und-Laxpair} (A_{\mu})_{ij} = (B_{\mu})_{ij} ~ e^{i(\tphi_{j}-\tphi_{i})} ~~~~~~ \textrm{with}
~~ \mu = 0,1 \eeq

\par{and we define}

\bea \nonumber & (B_{0})_{ij} & =  [\frac{3}{2}(r_{i}r_{j}'-r_{i}'r_{j}) +
\frac{3i}{2}r_{i}r_{j}(\tphi_{i}'+\tphi_{j}')-3ir_{i}r_{j}\sum_{k=1}^{3}r_{k}^{2}\tphi_{k}']~x  +
\frac{3i}{2}(3r_{i}r_{j}
-\delta_{ij})~x^{2}~,  \\
& & \\
& & \nonumber \\
& (B_{1})_{ij} & = i(3r_{i}r_{j} - \delta_{ij})~x ~. \eea

{}{}{}{}{}{}{}{}{}{}{}{}{}{}{}{}{}{}{}{}{}{}{}{}{}{}{}{}{}{}{}{}{}{}{}{}{}{}{}{}{}{}{}{}{}{}{}{}{}{}{}{}{}{}{}{}{}{}{}{}

\section{Strings in the $\gamma_{i}$-Deformed Background}

\subsection{Equations of Motion in the $\gamma_{i}$-Deformed Background}

\par{We now generalize the previous results to strings in the $\gamma_{i}$-deformed $\mathbb{R}\times S^{5}$
background, which are again rotating in three orthogonal planes.
The Lagrangian for this $\gamma_{i}$-deformed background in the
fast motion limit (derived in \cite{fro-roi-tsey2}) is}

\beq \cL = \sum_{i=1}^{3}{r_{i}^{2}\dphi_{i}} - \frac{\tlambda}{2}\{\sum_{i=1}^{3}(r_{i}')^{2} +
\sum_{^{i,j=1}_{i<j}}^{3}{r_{i}^{2}r_{j}^{2}(\phi_{i}'-\phi_{j}'-\sum\limits_{k=1}^{3}\epsilon_{ijk}\bgamma_{k})^{2}}
- \bgamma^{2}r_{1}^{2}r_{2}^{2}r_{3}^{2} + \Lambda (\sum_{i=1}^{3}r_{i}^{2} - 1) \} ~,\eeq

\par{where $\bgamma_{i} = \tgamma_{i}\cJ = \gamma_{i}J$ and $\bgamma = \sum\limits_{i=1}^{3}{\bgamma_{i}}$. }

\par{This $\gamma_{i}$-deformed Lagrangian (again re-defining $\tau \rightarrow \frac{1}{\tlambda}\tau$ and
$\cL \rightarrow -\frac{1}{\tlambda} \cL$) can also be written, using the constraint $\sum\limits_{i=1}^{3}{r_{i}^{2}}=1$, as}

\bea \label{eqn-def-Lag} \nonumber & \cL = & -\sum_{i=1}^{3}r_{i}^{2}\dphi_{i} + \frac{1}{2}
\sum_{i=1}^{3}(r_{i}')^{2} + \frac{1}{2}\sum_{^{i,j=1}_{i<j}}^{3}{r_{i}^{2}r_{j}^{2}
[(\phi_{i}'+\sum\limits_{l,m=1}^{3}\epsilon_{ilm}\bgamma_{l}r_{m}^{2}) - (\phi_{j}'+
\sum\limits_{l,m=1}^{3}\epsilon_{jlm}\bgamma_{l}r_{m}^{2})]^{2}} \\
& & + \frac{1}{2}\Lambda (\sum_{i=1}^{3}r_{i}^{2} - 1)~. \eea

\par{Varying the above Lagrangian with respect to $r_{i}$ and $\phi_{i}$ and then
using the constraint equation
$\sum\limits_{i=1}^{3}{r_{i}^{2}}=1$, which can be obtained by
varying with respect to $\Lambda$, gives the $\gamma_{i}$-deformed
equations of motion}

\bea \label{eqn-def-motion1} \nonumber & r_{i}'' = & -2r_{i}\{ \dphi_{i}
      + \sum_{l,m=1}^{3}{\epsilon_{ilm}r_{i}^{2}r_{l}^{2}\bgamma_{m}(\phi_{i}'-\phi_{l}'-\epsilon_{ilm}\bgamma_{m})}
      -\frac{1}{2}\sum_{l,m=1}^{3}{\epsilon_{ilm}r_{l}^{2}r_{m}^{2}(\bgamma_{l}+\bgamma_{m})
      (\phi_{l}'-\phi_{m}'-\epsilon_{ilm}\bgamma_{i})} \} \\
& & + r_{i}\sum_{k=1}^{3}{r_{k}^{2}[(\phi_{i}'+\sum\limits_{l,m=1}^{3}\epsilon_{ilm}\bgamma_{l}r_{m}^{2}) -
(\phi_{k}'+ \sum\limits_{l,m=1}^{3}\epsilon_{klm}\bgamma_{l}r_{m}^{2})]^{2}} + \Lambda r_{i}~, \\
\nonumber & & \\
\nonumber & & \\
\label{eqn-def-motion2} \nonumber & \dr_{i} = & \sum_{k=1}^{3}r_{k}(r_{i}r_{k})'[(\phi_{i}'+
\sum\limits_{l,m=1}^{3}\epsilon_{ilm}\bgamma_{l}r_{m}^{2})- (\phi_{k}'+
\sum\limits_{l,m=1}^{3}\epsilon_{klm}\bgamma_{l}r_{m}^{2})] \\
& & + \frac{1}{2}~r_{i}\sum_{k=1}^{3}{r_{k}^{2}[(\phi_{i}''+
2\sum\limits_{l,m=1}^{3}\epsilon_{ilm}\bgamma_{l}r_{m}r_{m}')- (\phi_{k}''+
2\sum\limits_{l,m=1}^{3}\epsilon_{klm}\bgamma_{l}r_{m}r_{m}')]}~. \eea

\bigskip

\par{Now, assuming this constraint is satisfied, the above equations of motion
(\ref{eqn-def-motion1}) and (\ref{eqn-def-motion2})
are equivalent to}

\bea \label{eqn-def_comm1} \nonumber  && r_{j}r_{i}''-r_{i}r_{j}''  = 2r_{i}r_{j}(\dphi_{j}-\dphi_{i}) \\
\nonumber  &&~~~~~ -2r_{i}r_{j} \{
\sum_{l,m=1}^{3}{\epsilon_{ilm}r_{i}^{2}r_{l}^{2}\bgamma_{m}(\phi_{i}'-\phi_{l}'-\epsilon_{ilm}\bgamma_{m})}
 -\frac{1}{2}\sum_{l,m=1}^{3}{\epsilon_{ilm}r_{l}^{2}r_{m}^{2}(\bgamma_{l}+\bgamma_{m})
      (\phi_{l}'-\phi_{m}'-\epsilon_{ilm}\bgamma_{i})} \} \\
\nonumber &&~~~~~ + 2r_{i}r_{j} \{
\sum_{l,m=1}^{3}{\epsilon_{jlm}r_{j}^{2}r_{l}^{2}\bgamma_{m}(\phi_{j}'-\phi_{l}'-\epsilon_{jlm}\bgamma_{m})}
-\frac{1}{2}\sum_{l,m=1}^{3}{\epsilon_{jlm}r_{l}^{2}r_{m}^{2}(\bgamma_{l}+\bgamma_{m})
      (\phi_{l}'-\phi_{m}'-\epsilon_{jlm}\bgamma_{j})} \} \\
\nonumber  &&~~~~~ + r_{i}r_{j}\sum_{k=1}^{3}{r_{k}^{2}[(\phi_{i}'+
\sum\limits_{l,m=1}^{3}\epsilon_{ilm}\bgamma_{l}r_{m}^{2})-(\phi_{k}'
+\sum\limits_{l,m=1}^{3}\epsilon_{klm}\bgamma_{l}r_{m}^{2})]^{2}} \\
 &&~~~~~ - r_{i}r_{j}\sum_{k=1}^{3}{r_{k}^{2}[(\phi_{j}'+\sum\limits_{l,m=1}^{3}\epsilon_{jlm}\bgamma_{l}r_{m}^{2})-
(\phi_{k}'+ \sum\limits_{l,m=1}^{3}\epsilon_{klm}\bgamma_{l}r_{m}^{2})]^{2}}~, \\
\nonumber & & \\
\nonumber & & \\
\label{eqn-def_comm2} \nonumber && \dr_{i}r_{j} + r_{i}\dr_{j}  =
r_{j}\sum_{k=1}^{3}{r_{k}(r_{i}r_{k})'[(\phi_{i}' +
\sum\limits_{l,m=1}^{3}\epsilon_{ilm}\bgamma_{l}r_{m}^{2}) - (\phi_{k}' + \sum\limits_{l,m=1}^{3}\epsilon_{klm}\bgamma_{l}r_{m}^{2})]} \\
\nonumber &&~~~~~ + r_{i}\sum_{k=1}^{3}{r_{k}(r_{j}r_{k})'[(\phi_{j}' +
\sum\limits_{l,m=1}^{3}\epsilon_{jlm}\bgamma_{l}r_{m}^{2}) -
(\phi_{k}' + \sum\limits_{l,m=1}^{3}\epsilon_{klm}\bgamma_{l}r_{m}^{2})]} \\
\nonumber &&~~~~~ + \frac{1}{2}r_{i}r_{j}\sum_{k=1}^{3}{r_{k}^{2}[(\phi_{i}''+
2\sum\limits_{l,m=1}^{3}\epsilon_{ilm}\bgamma_{l}r_{m}r_{m}')-(\phi_{k}''+2\sum\limits_{l,m=1}^{3}\epsilon_{klm}\bgamma_{l}r_{m}r_{m}')]} \\
 &&~~~~~ +
\frac{1}{2}r_{i}r_{j}\sum_{k=1}^{3}{r_{k}^{2}[(\phi_{j}''+
2\sum\limits_{l,m=1}^{3}\epsilon_{jlm}\bgamma_{l}r_{m}r_{m}')-(\phi_{k}''+2\sum\limits_{l,m=1}^{3}\epsilon_{klm}\bgamma_{l}r_{m}r_{m}')]}
~.  \eea

\subsection{Transformation from the Undeformed Equations of Motion to the $\gamma_{i}$-Deformed Equations of Motion}

\par{The transformation which takes the undeformed equations of motion into the
$\gamma_{i}$-deformed equations of motion is}

\bea \label{eqn-trans-phidot} \nonumber & \dtphi_{i} & = \dphi_{i} +
      \sum_{l,m=1}^{3}{\epsilon_{ilm}r_{i}^{2}r_{l}^{2}\bgamma_{m}(\phi_{i}'-\phi_{l}'-\epsilon_{ilm}\bgamma_{m})}
      -\frac{1}{2}\sum_{l,m=1}^{3}{\epsilon_{ilm}r_{l}^{2}r_{m}^{2}(\bgamma_{l}+\bgamma_{m})
      (\phi_{l}'-\phi_{m}'-\epsilon_{ilm}\bgamma_{i})}~, \\
    && \\
    \label{eqn-trans-phidash} & \tphi_{i}' & = \phi_{i}' + \sum\limits_{l,m=1}^{3}\epsilon_{ilm}\bgamma_{l}r_{m}^{2}~. \eea

\par{Now, for this transformation to be valid, we must have $(\dtphi_{i})' = (\tphi_{i}')\dot{}$.
Thus the compatibility condition, which must be satisfied, is}

\bea \label{eqn-def-compat} \nonumber && 2\sum\limits_{l,m=1}^{3}\epsilon_{ilm}\bgamma_{l}r_{m}\dr_{m} =
\p_{1}\{ \sum_{l,m=1}^{3}{\epsilon_{ilm}r_{i}^{2}r_{l}^{2}\bgamma_{m}(\phi_{i}'-\phi_{l}'-\epsilon_{ilm}\bgamma_{m})} \\
     & & ~~~~~~~~~~~~~~~~~~~~~~~~~~  -\frac{1}{2}\sum_{l,m=1}^{3}{\epsilon_{ilm}r_{l}^{2}r_{m}^{2}(\bgamma_{l}+\bgamma_{m})
      (\phi_{l}'-\phi_{m}'-\epsilon_{ilm}\bgamma_{i})} \} ~. \eea

\bigskip

\par{However, from the equation of motion (\ref{eqn-def-motion2}), we know that}

\beq  r_{i}\dr_{i} =
      \frac{1}{2}~\p_{1} \{ \sum_{k=1}^{3}{r_{i}^{2}r_{k}^{2}[(\phi_{i}'+ \sum\limits_{n,s=1}^{3}\epsilon_{ins}\bgamma_{n}r_{s}^{2})-
(\phi_{k}'+ \sum\limits_{n,s=1}^{3}\epsilon_{kns}\bgamma_{n}r_{s}^{2})]} \} \eeq

\par{and thus}

\beq \label{eqn-def-compat2} 2\sum\limits_{l,m=1}^{3}\epsilon_{ilm}\bgamma_{l}r_{m}\dr_{m} = \p_{1} \{
\sum_{k,l,m=1}^{3}{\epsilon_{ilm}\bgamma_{l}
      r_{m}^{2}r_{k}^{2}[(\phi_{m}'+ \sum\limits_{n,s=1}^{3}\epsilon_{mns}\bgamma_{n}r_{s}^{2})-
(\phi_{k}'+ \sum\limits_{n,s=1}^{3}\epsilon_{kns}\bgamma_{n}r_{s}^{2})]} \}~. \eeq

\bigskip

\par{By setting $i = 1,2$ and $3$, and evaluating equations (\ref{eqn-def-compat}) and (\ref{eqn-def-compat2})
 separately (see appendix B),  these equations can be shown to be the same. Thus the compatibility condition
is automatically satisfied if the $\gamma_{i}$-deformed equations
of motion (and the constraint equation) are valid.}

\subsection{Lax Pair in the $\gamma_{i}$-Deformed Background}

\par{The $\gamma_{i}$-deformed Lax pair shall now be derived from the undeformed one following a
similar procedure to that discussed in \cite{frolov}.}

\par{First the $\tphi_{i}$-dependence of the undeformed Lax pair will be gauged away.  The zero curvature condition is
$[\cD_{0},\cD_{1}] = 0$, where $\cD_{\mu} = \p_{\mu} - A_{\mu}$
with $\mu = 0,1$. This is equivalent to
$[M\cD_{0}M^{-1},M\cD_{1}M^{-1}] = 0$, for any invertible matrix
$M$, so we can change}

\beq \cD_{\mu} \rightarrow \ctD_{\mu} = M\cD_{\mu}M^{-1} =
\p_{\mu} + M\p_{\mu}M^{-1} - MA_{\mu}M^{-1}~. \eeq

\par{Thus an equivalent gauged Lax Pair is}

\beq \ctD_{\mu} = \p_{\mu} - \cR_{\mu}~, ~~~~ \textrm{where }
\cR_{\mu} = MA_{\mu}M^{-1} - M\p_{\mu}M^{-1}~. \eeq

\bigskip

\par{We shall take $M = ie^{i\tphi_{i}}\delta_{ij}$ and thus $M^{-1} = -ie^{-i\tphi_{i}}\delta_{ij}$. Therefore it
follows from equation  (\ref{eqn-und-Laxpair}) that the gauged undeformed Lax pair is $\ctD_{\mu} = \p_{\mu} -
\cR_{\mu}$, where}

\beq (\cR_{\mu})_{ij} = (B_{\mu})_{ij} + i\p_{\mu}\tphi_{i}~\delta_{ij} \eeq

\par{and thus, using the definition of $(B_{\mu})_{ij}$, we obtain }

\bea \nonumber & (\cR_{0})_{ij} & =  [\frac{3}{2}(r_{i}r_{j}'-r_{i}'r_{j}) +
\frac{3i}{2}r_{i}r_{j}(\tphi_{i}'+\tphi_{j}')-3ir_{i}r_{j}\sum_{k=1}^{3}r_{k}^{2}\tphi_{k}']~x +
\frac{3i}{2}(3r_{i}r_{j}
-\delta_{ij})~x^{2} + i\dtphi_{i}~\delta_{ij}~, \\
& & \\
& (\cR_{1})_{ij} & = i(3r_{i}r_{j} - \delta_{ij})~x
+i\tphi_{i}'~\delta_{ij}~. \eea

\bigskip

\par{We can now make use of the transformation (\ref{eqn-trans-phidot}) and (\ref{eqn-trans-phidash}) to obtain the
 gauged $\gamma_{i}$-deformed Lax pair $\ctD^{\gamma_{i}}_{\mu} = \p_{\mu} -
\cR^{\gamma_{i}}_{\mu}$, where}

\bea \nonumber && (\cR^{\gamma_{i}}_{0})_{ij}  = \frac{3}{2}(r_{i}r_{j}'-r_{i}'r_{j})~x +
\frac{3i}{2}r_{i}r_{j}[(\phi_{i}'+\sum\limits_{l,m=1}^{3}\epsilon_{ilm}\bgamma_{l}r_{m}^{2})
+ (\phi_{j}'+\sum\limits_{l,m=1}^{3}\epsilon_{jlm}\bgamma_{l}r_{m}^{2})]~x \\
\nonumber & &~~~ -3ir_{i}r_{j}\sum_{k=1}^{3}r_{k}^{2} (\phi_{k}' +
\sum\limits_{l,m=1}^{3}\epsilon_{klm}\bgamma_{l}r_{m}^{2})~x
 + \frac{3i}{2}(3r_{i}r_{j} -\delta_{ij})~x^{2}  \\
\nonumber & &~~~  + i\{\dphi_{i} +
      \sum_{l,m=1}^{3}{\epsilon_{ilm}r_{i}^{2}r_{l}^{2}\bgamma_{m}(\phi_{i}'-\phi_{l}'-\epsilon_{ilm}\bgamma_{m})}
-\frac{1}{2}\sum_{l,m=1}^{3}{\epsilon_{ilm}r_{l}^{2}r_{m}^{2}(\bgamma_{l}+\bgamma_{m})
      (\phi_{l}'-\phi_{m}'-\epsilon_{ilm}\bgamma_{i})} \}~\delta_{ij}~, \\
&&\\
&& (\cR^{\gamma_{i}}_{1})_{ij}  = i(3r_{i}r_{j} - \delta_{ij})~x +i(\phi_{i}' +
\sum\limits_{l,m=1}^{3}\epsilon_{ilm}\bgamma_{l}r_{m}^{2})~\delta_{ij} ~.\eea

\bigskip

\par{The zero curvature condition $[\ctD^{\gamma_{i}}_{0},\ctD^{\gamma_{i}}_{1}]=0$ is equivalent to}

\beq \p_{0}\cR^{\gamma_{i}}_{1} - \p_{1}\cR^{\gamma_{i}}_{0} - [\cR^{\gamma_{i}}_{0}, \cR^{\gamma_{i}}_{1}] = 0
\eeq

\bigskip \bigskip

\par{and the equations thus obtained from this gauged $\gamma_{i}$-deformed Lax pair
(see appendix C) are equations (\ref{eqn-def_comm1}) and
(\ref{eqn-def_comm2}), which are equivalent to the
$\gamma_{i}$-deformed equations of motion, and the compatibility
condition, which follows directly from these equations motion.}

\section{Conclusion}

\par{In this paper we have considered strings in $\mathbb{R}\times S^{5}$ (at the center of $AdS_{5}$) rotating
in three orthogonal planes in the non-supersymmetric $\gamma_{i}$-deformed background, which was constructed in
\cite{frolov} using a series of three TsT-transformations. Our starting point has been the string action in the
fast motion limit, in which the total angular momentum $J = J_{1}+J_{2}+J_{3}$ is large and we consider the
leading order in $\tlambda = \frac{\lambda}{J^{2}}$ (derived in \cite{fro-roi-tsey2}). This action is equivalent
on the gauge theory side to the effective action of the corresponding $\gamma_{i}$-deformed spin chain in the
continuum limit, in which the length of the spin chain becomes large \cite{fro-roi-tsey2}.}

\par{We have first reviewed the construction in \cite{fro-roi-tsey2,stef-tsey} of a
Lax pair in the undeformed case. It was
then demonstrated that there exists a consistent transformation
which takes the undeformed equations of motion into the
$\gamma_{i}$-deformed equations of motion.  This was used to
construct a Lax pair describing rotating strings in the the
$\gamma_{i}$-deformed background. Thus it was shown that the
$\gamma_{i}$-deformed theory remains integrable in the fast motion
limit.}

\par{A related topic for further investigation would be to attempt to calculate conserved quantities which
follow from this Lax pair.  Specifically one could try to
construct the monodromy matrix and thus the quasi-momenta as a
function of the spectral parameter for the undeformed and
$\gamma_{i}$-deformed theories in the fast motion limit. Another
interesting point of discussion is the physical meaning of the
transformation which was used to construct the
$\gamma_{i}$-deformed Lax pair.}

\section{Acknowledgments}

\par{I would like to thank my supervisor, Robert De Mello Koch, for suggesting this topic as an interesting
area of study and for his many helpful suggestions.  I would also like to acknowledge a National Research
Foundation Grant Holders Bursary.}

{}{}{}{}{}{}{}{}{}{}{}{}{}{}{}{}{}{}{}{}{}{}{}{}{}{}{}{}{}{}{}{}{}{}{}{}{}{}{}{}{}{}{}{}{}{}{}{}{}{}{}{}{}{}{}{}{}{}{}{}

\appendix

\section{Derivation of the Equations of Motion from the Lax Pair
in the Undeformed Background}

\subsection{Derivation of the Landau-Lifshitz Equation}

\par{The Lax pair is $\cD_{\mu} = \p_{\mu} - A_{\mu}$ \cite{fro-roi-tsey2}, where}

\bea & A_{0} & = \frac{1}{6}[N,\p_{1}N]x+\frac{3i}{2}Nx^{2}~, \\
     & A_{1} & = iNx ~. \eea

\par{The equation which must be satisfied is }

\beq \label{eqn-und-laxAeqn} \p_{0}A_{1} - \p_{1}A_{0} - [A_{0},
A_{1}] = 0~. \eeq

\par{Note also that $N$ satisfies the constraints $\textrm{Tr}(N)=0$ and $N^{2} = N + 2$
\cite{fro-roi-tsey2,stef-tsey}, due to the definition
$N_{ij} = 3U_{i}^{*}U_{j}-\delta_{ij}$, where $U_{i} =
r_{i}e^{i\tphi_{i}}$, and the constraint
$\sum\limits_{i=1}^{3}{r_{i}^{2}} = 1$}.

\par{Equation (\ref{eqn-und-laxAeqn}), written in terms of $N$, is }

\beq i\p_{0}Nx - \frac{1}{6}\p_{1}[N,\p_{1}N]x - \frac{3i}{2}\p_{1}Nx^{2} - \frac{i}{6}[[N,\p_{1}N], N]x^{2} = 0
\eeq

\par{and thus, equating different orders in $x$, }

\bea && \label{eqn-und-laxorderx1} i\p_{0}N = \frac{1}{6}\p_{1}[N,\p_{1}N]~, \\
     && \label{eqn-und-laxorderx2} \frac{3}{2}\p_{1}N = - \frac{1}{6}[[N,\p_{1}N], N] ~.\eea

\par{Equation (\ref{eqn-und-laxorderx2}) follows from the constraint $N^{2} = N + 2$
and equation (\ref{eqn-und-laxorderx1}) is equivalent to the
Landau-Lifshitz equation of motion \cite{fro-roi-tsey2,stef-tsey}
given by}

\beq \label{eqn-ll} i\p_{0}N = \frac{1}{6}[N,\p_{1}^{2}N]~. \eeq

\bigskip

\subsection{Derivation of the Undeformed Equations of Motion in
terms of $r_{i}$ and $\tphi_{i}$ from the Landau-Lifshitz
Equation}

\par{The definition of $N$ in component form is
$N_{ij} = 3r_{i}r_{j}e^{i(\tphi_{j}-\tphi_{i})}-\delta_{ij}$. Thus
we obtain}

\bea & \p_{0}N_{ij} & = 3[(\dr_{i}r_{j} + r_{i}\dr_{j}) +
ir_{i}r_{j}(\dtphi_{j}-\dtphi_{i})]~e^{i(\tphi_{j}-\tphi_{i})}~, \\
\nonumber && \\
& \p_{1}N_{ij} & = 3[(r_{i}'r_{j} + r_{i}r_{j}') +
ir_{i}r_{j}(\tphi_{j}'-\tphi_{i}')]~e^{i(\tphi_{j}-\tphi_{i})} \eea

\par{and hence}

\beq \p_{1}^{2}N_{ij} = 3[(r_{i}''r_{j} + 2r_{i}'r_{j}' + r_{i}r_{j}'') +
2i(r_{i}r_{j})'(\tphi_{j}'-\tphi_{i}')+ ir_{i}r_{j}(\tphi_{j}''-\tphi_{i}'') -
r_{i}r_{j}(\tphi_{j}'-\tphi_{i}')^{2}]~e^{i(\tphi_{j}-\tphi_{i})}~, \eeq

\bigskip

\par{from which it follows that}

\bea & [N, \p_{1}^{2}N]_{ij} & = \sum_{k=1}^{3}{N_{ik}\p_{1}^{2}N_{kj}} - \sum_{k=1}^{3}{\p_{1}^{2}N_{ik} N_{kj}} \\
                \nonumber    & & = 9\{ [r_{i}r_{j}'' - r_{j}r_{i}''] +
                            2i[r_{i}\sum_{k=1}^{3}{r_{k}(r_{k}r_{j})'(\tphi_{j}'-\tphi_{k}')} -
                            r_{j}\sum_{k=1}^{3}{r_{k}(r_{i}r_{k})'(\tphi_{k}'-\tphi_{i}')}] \\
                \nonumber & & ~~+ i[r_{i}r_{j}\sum_{k=1}^{3}{r_{k}^{2}(\tphi_{j}''-\tphi_{k}'')} -
                                r_{i}r_{j}\sum_{k=1}^{3}{r_{k}^{2}(\tphi_{k}''-\tphi_{i}'')}] \\
                & & ~~- [r_{i}r_{j}\sum_{k=1}^{3}{r_{k}^{2}(\tphi_{j}'-\tphi_{k}')^{2}}
                - r_{i}r_{j}\sum_{k=1}^{3}{r_{k}^{2}(\tphi_{k}'-\tphi_{i}')^{2}}]
                \}~e^{i(\tphi_{j}-\tphi_{i})} ~. \eea

\par{Now we use the Landau-Lifshitz equation (\ref{eqn-ll}) to obtain}

\bea \nonumber & &  i(\dr_{i}r_{j} + r_{i}\dr_{j}) - r_{i}r_{j}(\dtphi_{j}-\dtphi_{i}) \\
                \nonumber  & &    = \frac{1}{2}\{~ [r_{i}r_{j}'' - r_{j}r_{i}''] +
                            2i[r_{i}\sum_{k=1}^{3}{r_{k}(r_{k}r_{j})'(\tphi_{j}'-\tphi_{k}')} -
                            r_{j}\sum_{k=1}^{3}{r_{k}(r_{i}r_{k})'(\tphi_{k}'-\tphi_{i}')}] \\
                \nonumber & & ~~+ i[r_{i}r_{j}\sum_{k=1}^{3}{r_{k}^{2}(\tphi_{j}''-\tphi_{k}'')} -
                                r_{i}r_{j}\sum_{k=1}^{3}{r_{k}^{2}(\tphi_{k}''-\tphi_{i}'')}] \\
                 & & ~~- [r_{i}r_{j}\sum_{k=1}^{3}{r_{k}^{2}(\tphi_{j}'-\tphi_{k}')^{2}}
                - r_{i}r_{j}\sum_{k=1}^{3}{r_{k}^{2}(\tphi_{k}'-\tphi_{i}')^{2}}]~\}~.  \eea

\par{Therefore, equating the real and imaginary parts of the above equation, we find that }

\bea \nonumber \textrm{Re:} ~ &  r_{j}r_{i}'' - r_{i}r_{j}'' & = 2r_{i}r_{j}(\dtphi_{j}-\dtphi_{i}) +
r_{i}r_{j}\sum_{k=1}^{3}{r_{k}^{2}(\tphi_{i}' - \tphi_{k}')^{2}} - r_{i}r_{j}\sum_{k=1}^{3}{r_{k}^{2}(\tphi_{j}'
- \tphi_{k}')^{2}}~, \\
&& \\
\nonumber & & \\
\nonumber  \textrm{Im:} ~ &  \dr_{i}r_{j} + r_{i}\dr_{j} & =
r_{j}\sum_{k=1}^{3}{r_{k}(r_{i}r_{k})'(\tphi_{i}'-\tphi_{k}')} +
r_{i}\sum_{k=1}^{3}{r_{k}(r_{j}r_{k})'(\tphi_{j}'-\tphi_{k}')} \\
& & ~~+ \frac{1}{2}~r_{i}r_{j}\sum_{k=1}^{3}{r_{k}^{2}(\tphi_{i}''-\tphi_{k}'')} +
\frac{1}{2}~r_{i}r_{j}\sum_{k=1}^{3}{r_{k}^{2}(\tphi_{j}''-\tphi_{k}'')}~, \eea

\par{which can be compared with equations (\ref{eqn-und_comm1}) and (\ref{eqn-und_comm2}), and thus seen to be
equivalent to the undeformed equations of motion.}

{}{}{}{}{}{}{}{}{}{}{}{}{}{}{}{}{}{}{}{}{}{}{}{}{}{}{}{}{}{}{}{}{}{}{}{}{}{}{}{}{}{}{}{}{}{}{}{}{}{}{}{}{}{}{}{}{}{}{}{}

\section{Compatibility Condition for the
Transformation from the Undeformed Equations of Motion to the $\gamma_{i}$-Deformed Equations of Motion}

\par{The compatibility condition for the transformation is }

\bea \nonumber
      && 2\sum\limits_{l,m=1}^{3}\epsilon_{ilm}\bgamma_{l}r_{m}\dr_{m} =  \p_{1}\{
      \sum_{l,m=1}^{3}{\epsilon_{ilm}r_{i}^{2}r_{l}^{2}\bgamma_{m}(\phi_{i}'-\phi_{l}'-\epsilon_{ilm}\bgamma_{m})}\\
       & &~~~~~~~~~~~~~~~~~~~~~~~~~~ -\frac{1}{2}\sum_{l,m=1}^{3}{\epsilon_{ilm}r_{l}^{2}r_{m}^{2}(\bgamma_{l}+\bgamma_{m})
      (\phi_{l}'-\phi_{m}'-\epsilon_{ilm}\bgamma_{i})} \} ~.\eea

\bigskip

\par{The $\gamma_{i}$-deformed equation of motion (\ref{eqn-def-motion2}) gives }

\beq 2\sum_{l,m=1}^{3}\epsilon_{ilm}\bgamma_{l}r_{m}\dr_{m} = \p_{1} \{
\sum_{k,l,m=1}^{3}{\epsilon_{ilm}\bgamma_{l} r_{m}^{2}r_{k}^{2}[(\phi_{m}'+
\sum_{n,s=1}^{3}\epsilon_{mns}\bgamma_{n}r_{s}^{2})-(\phi_{k}'+
\sum_{n,s=1}^{3}\epsilon_{kns}\bgamma_{n}r_{s}^{2})]} \}~. \eeq

\bigskip

\par{First we shall evaulate }

\beq \{1\} \equiv \{
      \sum_{l,m=1}^{3}{\epsilon_{ilm}r_{i}^{2}r_{l}^{2}\bgamma_{m}(\phi_{i}'-\phi_{l}'-\epsilon_{ilm}\bgamma_{m})}
      -\frac{1}{2}\sum_{l,m=1}^{3}{\epsilon_{ilm}r_{l}^{2}r_{m}^{2}(\bgamma_{l}+\bgamma_{m})
      (\phi_{l}'-\phi_{m}'-\epsilon_{ilm}\bgamma_{i})} \} \eeq

\par{for $i = 1,2$ and $3$ as follows:}

\beq \{1\}_{i=1} = r_{1}^{2}r_{2}^{2}\bgamma_{3}(\phi_{1}'-\phi_{2}'-\bgamma_{3}) +
     r_{1}^{2}r_{3}^{2}\bgamma_{2}(\phi_{3}'-\phi_{1}'-\bgamma_{2})
     - r_{2}^{2}r_{3}^{2}(\bgamma_{2}+\bgamma_{3})(\phi_{2}'-\phi_{3}'-\bgamma_{1})~, \eeq

\beq \{1\}_{i=2} = r_{1}^{2}r_{2}^{2}\bgamma_{3}(\phi_{1}'-\phi_{2}'-\bgamma_{3}) +
     r_{2}^{2}r_{3}^{2}\bgamma_{1}(\phi_{2}'-\phi_{3}'-\bgamma_{1})
     - r_{1}^{2}r_{3}^{2}(\bgamma_{1}+\bgamma_{3})(\phi_{3}'-\phi_{1}'-\bgamma_{2})~, \eeq

\beq \{1\}_{i=3} = r_{1}^{2}r_{3}^{2}\bgamma_{2}(\phi_{3}'-\phi_{1}'-\bgamma_{2}) +
     r_{2}^{2}r_{3}^{2}\bgamma_{1}(\phi_{2}'-\phi_{3}'-\bgamma_{1})
     - r_{1}^{2}r_{2}^{2}(\bgamma_{1}+\bgamma_{2})(\phi_{1}'-\phi_{2}'-\bgamma_{3})~. \eeq

\bigskip

\par{Now we shall determine }

\beq \{2\} \equiv \{ \sum_{k,l,m=1}^{3}{\epsilon_{ilm}\bgamma_{l}
      r_{m}^{2}r_{k}^{2}[(\phi_{m}'+ \sum_{n,s=1}^{3}\epsilon_{mns}\bgamma_{n}r_{s}^{2})-(\phi_{k}'+
      \sum_{n,s=1}^{3}\epsilon_{kns}\bgamma_{n}r_{s}^{2})]} \} \eeq

\par{for $i = 1,2$ and $3$ as follows:}

\bea \nonumber & \{2\}_{i=1} & = \bgamma_{2}r_{3}^{2}(\phi_{3}' +
\bgamma_{1}r_{2}^{2}-\bgamma_{2}r_{1}^{2}) -
\bgamma_{3}r_{2}^{2}(\phi_{2}' +
\bgamma_{3}r_{1}^{2}-\bgamma_{1}r_{3}^{2}) -
\bgamma_{2}r_{3}^{2}(r_{1}^{2}\phi_{1}'+r_{2}^{2}\phi_{2}'+r_{3}^{2}\phi_{3}')\\
\nonumber & & ~~+\bgamma_{3}r_{2}^{2}(r_{1}^{2}\phi_{1}'+r_{2}^{2}\phi_{2}'+r_{3}^{2}\phi_{3}') \\
\nonumber & & \\
\nonumber & & = \bgamma_{2}r_{3}^{2}[(r_{1}^{2}+r_{2}^{2}+r_{3}^{2})\phi_{3}' +
\bgamma_{1}r_{2}^{2}-\bgamma_{2}r_{1}^{2}] - \bgamma_{3}r_{2}^{2}[(r_{1}^{2}+r_{2}^{2}+r_{3}^{2})\phi_{2}' +
\bgamma_{3}r_{1}^{2}-\bgamma_{1}r_{3}^{2}] \\
\nonumber & & ~~-\bgamma_{2}r_{3}^{2}(r_{1}^{2}\phi_{1}'+r_{2}^{2}\phi_{2}'+r_{3}^{2}\phi_{3}') +
\bgamma_{3}r_{2}^{2}(r_{1}^{2}\phi_{1}'+r_{2}^{2}\phi_{2}'+r_{3}^{2}\phi_{3}') \\
\nonumber & & \\
\nonumber & & = r_{1}^{2}r_{2}^{2}\bgamma_{3}(\phi_{1}'-\phi_{2}'
- \bgamma_{3})
        + r_{1}^{2}r_{3}^{2}\bgamma_{2}(\phi_{3}'-\phi_{1}' - \bgamma_{2})
        - r_{2}^{2}r_{3}^{2}(\bgamma_{2}+\bgamma_{3})(\phi_{2}'-\phi_{3}' - \bgamma_{1})~, \\
& & \eea

\bea \nonumber & \{2\}_{i=2} & = \bgamma_{3}r_{1}^{2}(\phi_{1}' +
\bgamma_{2}r_{3}^{2}-\bgamma_{3}r_{2}^{2}) -
\bgamma_{1}r_{3}^{2}(\phi_{3}' +
\bgamma_{1}r_{2}^{2}-\bgamma_{2}r_{1}^{2}) -
\bgamma_{3}r_{1}^{2}(r_{1}^{2}\phi_{1}'+r_{2}^{2}\phi_{2}'+r_{3}^{2}\phi_{3}')\\
\nonumber & & ~~+\bgamma_{1}r_{3}^{2}(r_{1}^{2}\phi_{1}'+r_{2}^{2}\phi_{2}'+r_{3}^{2}\phi_{3}') \\
\nonumber & & \\
\nonumber & & = \bgamma_{3}r_{1}^{2}[(r_{1}^{2}+r_{2}^{2}+r_{3}^{2})\phi_{1}' +
\bgamma_{2}r_{3}^{2}-\bgamma_{3}r_{2}^{2}] - \bgamma_{1}r_{3}^{2}[(r_{1}^{2}+r_{2}^{2}+r_{3}^{2})\phi_{3}' +
\bgamma_{1}r_{2}^{2}-\bgamma_{2}r_{1}^{2}] \\
\nonumber & & ~~-\bgamma_{3}r_{1}^{2}(r_{1}^{2}\phi_{1}'+r_{2}^{2}\phi_{2}'+r_{3}^{2}\phi_{3}') +
\bgamma_{1}r_{3}^{2}(r_{1}^{2}\phi_{1}'+r_{2}^{2}\phi_{2}'+r_{3}^{2}\phi_{3}') \\
\nonumber & & \\
\nonumber & & = r_{1}^{2}r_{2}^{2}\bgamma_{3}(\phi_{1}'-\phi_{2}'
- \bgamma_{3})
        + r_{2}^{2}r_{3}^{2}\bgamma_{1}(\phi_{2}'-\phi_{3}' - \bgamma_{1})
        - r_{1}^{2}r_{3}^{2}(\bgamma_{1}+\bgamma_{3})(\phi_{3}'-\phi_{1}' - \bgamma_{2})~, \\
& &  \eea

\bea \nonumber & \{2\}_{i=3} & = \bgamma_{1}r_{2}^{2}(\phi_{2}' +
\bgamma_{3}r_{1}^{2}-\bgamma_{1}r_{3}^{2}) -
\bgamma_{2}r_{1}^{2}(\phi_{1}' +
\bgamma_{2}r_{3}^{2}-\bgamma_{3}r_{2}^{2}) -
\bgamma_{1}r_{2}^{2}(r_{1}^{2}\phi_{1}'+r_{2}^{2}\phi_{2}'+r_{3}^{2}\phi_{3}')
\\
\nonumber & & ~~+\bgamma_{2}r_{1}^{2}(r_{1}^{2}\phi_{1}'+r_{2}^{2}\phi_{2}'+r_{3}^{2}\phi_{3}') \\
\nonumber & & \\
\nonumber & & = \bgamma_{1}r_{2}^{2}[(r_{1}^{2}+r_{2}^{2}+r_{3}^{2})\phi_{2}' +
\bgamma_{3}r_{1}^{2}-\bgamma_{1}r_{3}^{2}] - \bgamma_{2}r_{1}^{2}[(r_{1}^{2}+r_{2}^{2}+r_{3}^{2})\phi_{1}' +
\bgamma_{2}r_{3}^{2}-\bgamma_{3}r_{2}^{2}] \\
\nonumber & & ~~- \bgamma_{1}r_{2}^{2}(r_{1}^{2}\phi_{1}'+r_{2}^{2}\phi_{2}'+r_{3}^{2}\phi_{3}') +
\bgamma_{2}r_{1}^{2}(r_{1}^{2}\phi_{1}'+r_{2}^{2}\phi_{2}'+r_{3}^{2}\phi_{3}') \\
\nonumber & & \\
\nonumber & & = r_{1}^{2}r_{3}^{2}\bgamma_{2}(\phi_{3}'-\phi_{1}'
- \bgamma_{2}) + r_{2}^{2}r_{3}^{2}\bgamma_{1}(\phi_{2}'-\phi_{3}'
- \bgamma_{1})
- r_{1}^{2}r_{2}^{2}(\bgamma_{1}+\bgamma_{2})(\phi_{1}'-\phi_{2}' - \bgamma_{3}) \\
& & \eea

\par{using the constraint $\sum\limits_{i=1}^{3}{r_{i}^{2}}=1$. We have thus show that
$\{1\} = \{2\}$ and therefore the compatibility condition follows
from the $\gamma_{i}$-deformed equations of motion.}

{}{}{}{}{}{}{}{}{}{}{}{}{}{}{}{}{}{}{}{}{}{}{}{}{}{}{}{}{}{}{}{}{}{}{}{}{}{}{}{}{}{}{}{}{}{}{}{}{}{}{}{}{}{}{}{}{}{}{}{}

\section{Derivation of the Equations of Motion from the Gauged
Lax Pair in the $\gamma_{i}$-Deformed Background}

\par{The gauged $\gamma_{i}$-deformed Lax pair is $\ctD^{\gamma_{i}}_{\mu} = \p_{\mu} - \cR^{\gamma_{i}}_{\mu}$, where}

\bea \label{eqn-def-Laxpair0-app} \nonumber && (\cR^{\gamma_{i}}_{0})_{ij} =
\frac{3}{2}(r_{i}r_{j}'-r_{i}'r_{j})~x + \frac{3i}{2}r_{i}r_{j}[(\phi_{i}' +
\sum_{l,m=1}^{3}\epsilon_{ilm}\bgamma_{l}r_{m}^{2}) + (\phi_{j}' +
\sum_{l,m=1}^{3}\epsilon_{jlm}\bgamma_{l}r_{m}^{2})]~x \\
\nonumber &&~~~ -3ir_{i}r_{j}\sum_{k=1}^{3}r_{k}^{2} (\phi_{k}' +
\sum_{l,m=1}^{3}\epsilon_{klm}\bgamma_{l}r_{m}^{2})~x
 + \frac{3i}{2}(3r_{i}r_{j} -\delta_{ij})~x^{2}  \\
\nonumber&&~~~  + i\{ \dphi_{i} +
      \sum_{l,m=1}^{3}{\epsilon_{ilm}r_{i}^{2}r_{l}^{2}\bgamma_{m}(\phi_{i}'-\phi_{l}'-\epsilon_{ilm}\bgamma_{m})}
-\frac{1}{2}\sum_{l,m=1}^{3}{\epsilon_{ilm}r_{l}^{2}r_{m}^{2}(\bgamma_{l}+\bgamma_{m})
      (\phi_{l}'-\phi_{m}'-\epsilon_{ilm}\bgamma_{i})} \}~\delta_{ij}~, \\
 & & \\
\label{eqn-def-Laxpair1-app} && (\cR^{\gamma_{i}}_{1})_{ij} = i(3r_{i}r_{j} - \delta_{ij})~x +i(\phi_{i}' +
\sum_{l,m=1}^{3}\epsilon_{ilm}\bgamma_{l}r_{m}^{2})~\delta_{ij}~. \eea

\bigskip

\par{The zero curvature condition

\beq \p_{0}\cR^{\gamma_{i}}_{1} - \p_{1}\cR^{\gamma_{i}}_{0} - [\cR^{\gamma_{i}}_{0}, \cR^{\gamma_{i}}_{1}] = 0
\eeq

\par{must now be satisfied.}

\bigskip

\par{We substitute equations (\ref{eqn-def-Laxpair0-app}) and (\ref{eqn-def-Laxpair1-app}) into this condition
and equate different orders of the spectral parameter $x$ as
follows: }

\bigskip

\par{$\underline{~O(x^{0}):}$ ~~ At zeroth order in the spectral parameter we obtain}

\bea\nonumber &&i\p_{0}(\phi_{i}'+\sum_{l,m=1}^{3}\epsilon_{ilm}\bgamma_{l}r_{m}^{2})~\delta_{ij}
- i\p_{1}\{ \dphi_{i}+\sum_{l,m=1}^{3}{\epsilon_{ilm}r_{i}^{2}r_{l}^{2}\bgamma_{m}(\phi_{i}'-\phi_{l}'-\epsilon_{ilm}\bgamma_{m})}\\
&&-\frac{1}{2}\sum_{l,m=1}^{3}{\epsilon_{ilm}r_{l}^{2}r_{m}^{2}(\bgamma_{l}+\bgamma_{m})
      (\phi_{l}'-\phi_{m}'-\epsilon_{ilm}\bgamma_{i})} \}~\delta_{ij} = 0 \eea

\par{and therefore}

\bea\nonumber&&\p_{0}(\sum_{l,m=1}^{3}\epsilon_{ilm}\bgamma_{l}r_{m}^{2})
= \p_{1}\{\sum_{l,m=1}^{3}{\epsilon_{ilm}r_{i}^{2}r_{l}^{2}\bgamma_{m}(\phi_{i}'-\phi_{l}'-\epsilon_{ilm}\bgamma_{m})}\\
&& ~~~~~~~~~~~~~~~~~~~~~~~~~~
-\frac{1}{2}\sum_{l,m=1}^{3}{\epsilon_{ilm}r_{l}^{2}r_{m}^{2}(\bgamma_{l}+\bgamma_{m})
      (\phi_{l}'-\phi_{m}'-\epsilon_{ilm}\bgamma_{i})} \} ~. \eea

\par{This is just the compatibility condition for the transformation from the undeformed equations of motion
to the $\gamma_{i}$-deformed equations of motion.}

\bigskip

\par{$\underline{~O(x^{1}):}$ ~~ At first order in the spectral parameter we find that}

\bea \nonumber & & 3i(\dr_{i}r_{j} + r_{i}\dr_{j}) - \frac{3}{2}(r_{i}r_{j}''-r_{i}''r_{j})
-\frac{3i}{2}\p_{1}\{ r_{i}r_{j}[(\phi_{i}' + \sum_{l,m=1}^{3}\epsilon_{ilm}\bgamma_{l}r_{m}^{2}) + (\phi_{j}' +
\sum_{l,m=1}^{3}\epsilon_{jlm}\bgamma_{l}r_{m}^{2})] \}  \\
\nonumber & & + 3i\p_{1}\{r_{i}r_{j}\sum_{k=1}^{3}{r_{k}^{2}(\phi_{k}' +
\sum_{l,m=1}^{3}\epsilon_{klm}\bgamma_{l}r_{m}^{2})}\}
 + 3r_{i}r_{j}(\dphi_{i}-\dphi_{j}) \\
\nonumber & & + 3r_{i}r_{j} \{
\sum_{l,m=1}^{3}{\epsilon_{ilm}r_{i}^{2}r_{l}^{2}\bgamma_{m}(\phi_{i}'-\phi_{l}'-\epsilon_{ilm}\bgamma_{m})}
      -\frac{1}{2}\sum_{l,m=1}^{3}{\epsilon_{ilm}r_{l}^{2}r_{m}^{2}(\bgamma_{l}+\bgamma_{m})
      (\phi_{l}'-\phi_{m}'-\epsilon_{ilm}\bgamma_{i})} \} \\
\nonumber & & - 3r_{i}r_{j} \{
\sum_{l,m=1}^{3}{\epsilon_{jlm}r_{j}^{2}r_{l}^{2}\bgamma_{m}(\phi_{j}'-\phi_{l}'-\epsilon_{jlm}\bgamma_{m})}
      -\frac{1}{2}\sum_{l,m=1}^{3}{\epsilon_{jlm}r_{l}^{2}r_{m}^{2}(\bgamma_{l}+\bgamma_{m})
      (\phi_{l}'-\phi_{m}'-\epsilon_{jlm}\bgamma_{j})} \} \\
\nonumber & & + \frac{3i}{2}(r_{i}r_{j}'-r_{i}'r_{j})[(\phi_{i}' +
\sum_{l,m=1}^{3}\epsilon_{ilm}\bgamma_{l}r_{m}^{2}) - (\phi_{j}' + \sum_{l,m=1}^{3}\epsilon_{jlm}\bgamma_{l}r_{m}^{2})] \\
\nonumber & & -~\frac{3}{2}r_{i}r_{j}[(\phi_{i}' + \sum_{l,m=1}^{3}\epsilon_{ilm}\bgamma_{l}r_{m}^{2}) +
(\phi_{j}' + \sum_{l,m=1}^{3}\epsilon_{jlm}\bgamma_{l}r_{m}^{2})][(\phi_{i}' +
\sum_{n,s=1}^{3}\epsilon_{ins}\bgamma_{n}r_{s}^{2}) - (\phi_{j}' +
\sum_{n,s=1}^{3}\epsilon_{jns}\bgamma_{n}r_{s}^{2})] \\
& & +~ 3r_{i}r_{j}\sum_{k=1}^{3}{r_{k}^{2}(\phi_{k}' +
\sum_{l,m=1}^{3}\epsilon_{klm}\bgamma_{l}r_{m}^{2})[(\phi_{i}' +
\sum_{n,s=1}^{3}\epsilon_{ins}\bgamma_{n}r_{s}^{2}) - (\phi_{j}' +
\sum_{n,s=1}^{3}\epsilon_{jns}\bgamma_{n}r_{s}^{2})]} = 0 ~. \eea

\par{Thus, equating the real and imaginary parts, we obtain}

\bea \label{eqn-def-mot-Re}~ \nonumber \textrm{Re:} &&
r_{i}''r_{j}-r_{i}r_{j}''  =
2r_{i}r_{j}(\dphi_{j}-\dphi_{i}) \\
\nonumber & &~~ - 2r_{i}r_{j} \{
\sum_{l,m=1}^{3}{\epsilon_{ilm}r_{i}^{2}r_{l}^{2}\bgamma_{m}(\phi_{i}'-\phi_{l}'-\epsilon_{ilm}\bgamma_{m})}
-\frac{1}{2}\sum_{l,m=1}^{3}{\epsilon_{ilm}r_{l}^{2}r_{m}^{2}(\bgamma_{l}+\bgamma_{m})
      (\phi_{l}'-\phi_{m}'-\epsilon_{ilm}\bgamma_{i})} \} \\
\nonumber & &~~ + 2r_{i}r_{j} \{
\sum_{l,m=1}^{3}{\epsilon_{jlm}r_{j}^{2}r_{l}^{2}\bgamma_{m}(\phi_{j}'-\phi_{l}'-\epsilon_{jlm}\bgamma_{m})}
-\frac{1}{2}\sum_{l,m=1}^{3}{\epsilon_{jlm}r_{l}^{2}r_{m}^{2}(\bgamma_{l}+\bgamma_{m})
      (\phi_{l}'-\phi_{m}'-\epsilon_{jlm}\bgamma_{j})} \} \\
\nonumber & &~~ + r_{i}r_{j}[(\phi_{i}' +
\sum_{l,m=1}^{3}\epsilon_{ilm}\bgamma_{l}r_{m}^{2})^{2} - (\phi_{j}' + \sum_{l,m=1}^{3}\epsilon_{jlm}\bgamma_{l}r_{m}^{2})^{2}]\\
\nonumber & &~~
-2r_{i}r_{j}\sum_{k=1}^{3}{r_{k}^{2}(\phi_{k}'+\sum_{l,m=1}^{3}\epsilon_{klm}\bgamma_{l}r_{m}^{2})[(\phi_{i}'
+\sum_{n,s=1}^{3}\epsilon_{ins}\bgamma_{n}r_{s}^{2}) - (\phi_{j}' +
\sum_{n,s=1}^{3}\epsilon_{jns}\bgamma_{n}r_{s}^{2})]}~,\\
 & & \\
\nonumber & & \\
\label{eqn-def-mot-Im}~ \nonumber \textrm{Im:} && \dr_{i}r_{j} + r_{i}\dr_{j} = \frac{1}{2}\p_{1}\{
r_{i}r_{j}[(\phi_{i}'+\sum_{l,m=1}^{3}\epsilon_{ilm}\bgamma_{l}r_{m}^{2})+(\phi_{j}'
+ \sum_{l,m=1}^{3}\epsilon_{jlm}\bgamma_{l}r_{m}^{2})] \} \\
\nonumber & &~~~~~~~~ -
\p_{1}\{r_{i}r_{j}\sum_{k=1}^{3}{r_{k}^{2}(\phi_{k}' +
\sum_{l,m=1}^{3}\epsilon_{klm}\bgamma_{l}r_{m}^{2})}\} \\
& & ~~~~~~~~ - \frac{1}{2}(r_{i}r_{j}'-r_{i}'r_{j})[(\phi_{i}' +
\sum_{l,m=1}^{3}\epsilon_{ilm}\bgamma_{l}r_{m}^{2}) - (\phi_{j}' +
\sum_{l,m=1}^{3}\epsilon_{jlm}\bgamma_{l}r_{m}^{2})]~.\eea

\bigskip

\par{Now equation (\ref{eqn-def-mot-Re}) is equivalent to }

\bea \label{eqn-def-lax-comm1} \nonumber && r_{j}r_{i}''-r_{i}r_{j}'' = 2r_{i}r_{j}(\dphi_{j}-\dphi_{i}) \\
\nonumber & &~~~~ - 2r_{i}r_{j} \{
\sum_{l,m=1}^{3}{\epsilon_{ilm}r_{i}^{2}r_{l}^{2}\bgamma_{m}(\phi_{i}'-\phi_{l}'-\epsilon_{ilm}\bgamma_{m})}
-\frac{1}{2}\sum_{l,m=1}^{3}{\epsilon_{ilm}r_{l}^{2}r_{m}^{2}(\bgamma_{l}+\bgamma_{m})
      (\phi_{l}'-\phi_{m}'-\epsilon_{ilm}\bgamma_{i})} \} \\
\nonumber & &~~~~ + 2r_{i}r_{j} \{
\sum_{l,m=1}^{3}{\epsilon_{jlm}r_{j}^{2}r_{l}^{2}\bgamma_{m}(\phi_{j}'-\phi_{l}'-\epsilon_{jlm}\bgamma_{m})}
-\frac{1}{2}\sum_{l,m=1}^{3}{\epsilon_{jlm}r_{l}^{2}r_{m}^{2}(\bgamma_{l}+\bgamma_{m})
      (\phi_{l}'-\phi_{m}'-\epsilon_{jlm}\bgamma_{j})} \} \\
\nonumber & &~~~~ + r_{i}r_{j}\sum_{k=1}^{3}{r_{k}^{2}[(\phi_{i}'+
\sum_{l,m=1}^{3}\epsilon_{ilm}\bgamma_{l}r_{m}^{2})-(\phi_{k}' +
\sum_{l,m=1}^{3}\epsilon_{klm}\bgamma_{l}r_{m}^{2})]^{2}} \\
& &~~~~ -r_{i}r_{j}\sum_{k=1}^{3}{r_{k}^{2}[(\phi_{j}'+\sum_{l,m=1}^{3}\epsilon_{jlm}\bgamma_{l}r_{m}^{2})-
(\phi_{k}'+ \sum_{l,m=1}^{3}\epsilon_{klm}\bgamma_{l}r_{m}^{2})]^{2}}~, \eea

\par{which can be seen by multiplying out the last two squared terms and noting that
$\sum\limits_{i=1}^{3}{r_{i}^{2}}=1$.}

\par{Equation (\ref{eqn-def-mot-Im}) can be written as}

\bea \nonumber && \dr_{i}r_{j} + r_{i}\dr_{j}  = \frac{1}{2}\ (r_{i}r_{j}' + r_{i}'r_{j})[\phi_{i}'+
\sum_{l,m=1}^{3}\epsilon_{ilm}\bgamma_{l}r_{m}^{2}) + (\phi_{j}' +
\sum_{l,m=1}^{3}\epsilon_{jlm}\bgamma_{l}r_{m}^{2})]  \\
\nonumber & &~~~~~~ + \frac{1}{2} r_{i}r_{j}[(\phi_{i}''+2\sum_{l,m=1}^{3}\epsilon_{ilm}\bgamma_{l}r_{m}r_{m}')
+(\phi_{j}''+2\sum_{l,m=1}^{3}\epsilon_{jlm}\bgamma_{l}r_{m}r_{m}')]  \\
\nonumber &&~~~~~~- (r_{i}r_{j}'+r_{i}'r_{j})\sum_{k=1}^{3}{r_{k}^{2}(\phi_{k}' +
\sum_{l,m=1}^{3}\epsilon_{klm}\bgamma_{l}r_{m}^{2})} -2r_{i}r_{j}\sum_{k=1}^{3}{r_{k}r_{k}'(\phi_{k}' +
\sum_{l,m=1}^{3}\epsilon_{klm}\bgamma_{l}r_{m}^{2})} \\
\nonumber & &~~~~~~ -r_{i}r_{j}\sum_{k=1}^{3}{r_{k}^{2}(\phi_{k}''
+2\sum_{l,m=1}^{3}\epsilon_{klm}\bgamma_{l}r_{m}r_{m}')}\\
\nonumber&&~~~~~~ - \frac{1}{2}(r_{i}r_{j}'-r_{i}'r_{j})[(\phi_{i}' +
\sum_{l,m=1}^{3}\epsilon_{ilm}\bgamma_{l}r_{m}^{2})
- (\phi_{j}'+\sum_{l,m=1}^{3}\epsilon_{jlm}\bgamma_{l}r_{m}^{2})] \\
\nonumber && \\
\nonumber && \\
\nonumber & &  = r_{i}'r_{j}(\phi_{i}'+\sum_{l,m=1}^{3}\epsilon_{ilm}\bgamma_{l}r_{m}^{2}) +
r_{i}r_{j}'(\phi_{j}'+\sum_{l,m=1}^{3}\epsilon_{jlm}\bgamma_{l}r_{m}^{2})\\
\nonumber&&~~~~~-2r_{i}r_{j}\sum_{k=1}^{3}{r_{k}r_{k}'(\phi_{k}'+\sum_{l,m=1}^{3}\epsilon_{klm}\bgamma_{l}r_{m}^{2})}
-r_{i}'r_{j}\sum_{k=1}^{3}{r_{k}^{2}(\phi_{k}'+\sum_{l,m=1}^{3}\epsilon_{klm}\bgamma_{l}r_{m}^{2})}\\
\nonumber&&~~~~~
-r_{i}r_{j}'\sum_{k=1}^{3}{r_{k}^{2}(\phi_{k}'+\sum_{l,m=1}^{3}\epsilon_{klm}\bgamma_{l}r_{m}^{2})}
 + \frac{1}{2}r_{i}r_{j}(\phi_{i}'' +
2\sum_{l,m=1}^{3}\epsilon_{ilm}\bgamma_{l}r_{m}r_{m}')\\
&&~~~~~ + \frac{1}{2}r_{i}r_{j}(\phi_{j}'' + 2\sum_{l,m=1}^{3}\epsilon_{jlm}\bgamma_{l}r_{m}r_{m}')
-r_{i}r_{j}\sum_{k=1}^{3}{r_{k}^{2}(\phi_{k}''+2\sum_{l,m=1}^{3}\epsilon_{klm}\bgamma_{l}r_{m}r_{m}')}~, \eea

\par{which implies, if one uses
the constraint $\sum\limits_{i=1}^{3}{r_{i}^{2}} = 1$ and thus $\sum\limits_{i=1}^{3}{r_{i}r_{i}'} = 0$, that}

\bea \label{eqn-def-lax-comm2} \nonumber && \dr_{i}r_{j} +
r_{i}\dr_{j} = r_{j}\sum_{k=1}^{3}{r_{k}(r_{i}r_{k})'[(\phi_{i}' +
\sum_{l,m=1}^{3}\epsilon_{ilm}\bgamma_{l}r_{m}^{2}) - (\phi_{k}' + \sum_{l,m=1}^{3}\epsilon_{klm}\bgamma_{l}r_{m}^{2})]} \\
\nonumber &&~~~~+ r_{i}\sum_{k=1}^{3}{r_{k}(r_{j}r_{k})'[(\phi_{j}' +
\sum_{l,m=1}^{3}\epsilon_{jlm}\bgamma_{l}r_{m}^{2}) -
(\phi_{k}' + \sum_{l,m=1}^{3}\epsilon_{klm}\bgamma_{l}r_{m}^{2})]} \\
\nonumber & &~~~~ +
\frac{1}{2}r_{i}r_{j}\sum_{k=1}^{3}{r_{k}^{2}[(\phi_{i}''+
2\sum_{l,m=1}^{3}\epsilon_{ilm}\bgamma_{l}r_{m}r_{m}')-(\phi_{k}''+2\sum_{l,m=1}^{3}\epsilon_{klm}\bgamma_{l}r_{m}r_{m}')]} \\
 &&~~~~ +
\frac{1}{2}r_{i}r_{j}\sum_{k=1}^{3}{r_{k}^{2}[(\phi_{j}''+
2\sum_{l,m=1}^{3}\epsilon_{jlm}\bgamma_{l}r_{m}r_{m}')-(\phi_{k}''+2\sum_{l,m=1}^{3}\epsilon_{klm}\bgamma_{l}r_{m}r_{m}')]}~.
\eea

\par{Equations (\ref{eqn-def-lax-comm1}) and (\ref{eqn-def-lax-comm2}) are the same as equations
(\ref{eqn-def_comm1}) and (\ref{eqn-def_comm2}), and are thus equivalent to the $\gamma_{i}$-deformed equations
of motion.}

\bigskip

\par{$\underline{~O(x^{2}):}$ ~~ At second order in the spectral parameter, one obtains an equation
which is trivially satisfied, again using the constraint
$\sum\limits_{i=1}^{3}{r_{i}^{2}} = 1$ and hence that
$\sum\limits_{i=1}^{3}{r_{i}r_{i}'} = 0$.}

\end{document}